\documentclass[aps,pre,twocolumn]{revtex4-2}

\usepackage{amsmath}
\usepackage{amsfonts}
\usepackage{amssymb}
\usepackage{graphicx}

\newcommand{\prs}[1]{{\left(#1\right)}}
\newcommand{\col}[1]{{\left[#1\right]}}

\newcommand{\probd}[1]{{\mathcal{P}_d\prs{#1}}}

\newcommand{\TS}{{\mathcal{S}}}
\newcommand{\cut}[1]{}

\begin{document}

\title{Unveiling the Link between Complexity and Symmetry: Statistical Asymmetry}
\author{Alamino, R.C.}
\affiliation{Applied AI and Robotics Department, Aston University, Birmingham, B29 6RP, United Kingdom}

\begin{abstract}
The concept of complexity appears in virtually all areas of knowledge. Its intuitive meaning shares similarities across fields, but disagreements between its details hinders a general definition, leading to a plethora of proposed measurements. While each might be appropriated to the problems it addresses, the lack of an underlying fundamental principle prevents the development of a unified theory. Here it is shown that the statistics of the amount of symmetry broken by systems can be used as such unifying principle. A general methodology is outlined and explicit expressions are given for cases in which it can capture the behavior of the two main groups of complexities currently in use. The presented results demonstrate that statistical asymmetry is an appropriate foundation for characterizing the general concept of complexity.  
\end{abstract}

\maketitle

\section{Introduction}

\textit{Symmetry} is one of the most important and ubiquitous concepts in science. In physics, gauge symmetry is the underlying principle of quantum field theory, while special and general relativity are based on Lorentz and diffeomorphism symmetries respectively. The symmetries of the physical action lead directly to conservation laws through Noether's theorem. On the other hand, its breaking can lead to spectacular consequences such as the generation of mass through the Higgs mechanism. Its role in other disciplines is no less important. In chemistry, it is a tool for classification of crystals and characterization of behavior in molecules \cite{Jaffe02}; in biology, while evolution seems to favour symmetry \cite{Johnston22}, several successful adaptations of organisms are linked to its loss \cite{Cubas01}, and its role in art and aesthetics is a continuous topic of study \cite{Easton23}.

An object's symmetry can be characterized by identifying a set of transformations acting on it which leaves some object-related quantity invariant, in which case the object is said to be symmetric under the action of the set. Such sets usually have a group structure and are said to be the symmetry groups of the object. The nature of the object is largely general -- it can be anything from an actual physical system to an abstract pattern.

Early studies in psychology have demonstrated that symmetry properties of geometric patterns facilitate memorization and recognition \cite{Attneave55}. Inspired by those results, continuous measures of the amount of approximate symmetry contained in a pattern under a group of transformations have been proposed. Some examples are Yodogawa's symmetropy \cite{Yodogawa82} and the symmetry distance (SD) proposed by Zabrodsky \textit{et al.} \cite{Zabrodsky95}.

Another series of psychology experiments during the 70's have shown that symmetry is commonly associated to the perceived intuitive idea of complexity of visual patterns \cite{Chipman77}. Like symmetry, the concept of complexity can be found in all scientific disciplines, although with related but nonequivalent meanings. The complexity of organisms, for instance, is a central concept in evolutionary biology \cite{Taylor16}. Proposed definitions usually rank objects or systems as diverse as polymer chains \cite{Matsubara18}, emergent physical properties \cite{Bar97}, quantum fields \cite{Camargo19}, Boolean networks \cite{Correale06} or binary sequences \cite{Li97}, according to features which are relevant for each particular field and applications therein. Some features are shared between such definitions, but there are also many points of disagreement. Finding a general framework capable of unifying the existing points of view or whether that is even possible is an open problem, which is evident by the large, ever growing number of complexity measures in the literature \cite{Wackerbauer94, Allouche12}. Such a framework could shed light on the fundamental meaning of complexity and also become an important tool in applications such as time series \cite{Rosso07, Politi17}, image analysis \cite{Zanette18}, design of nanostructures \cite{Arapis22} or the detection of adverse health conditions \cite{Silva2012, Frost18}. 

Formal mathematical connections between complexity and symmetry have been analyzed before \cite{Baake13, Alamino15}, but no general theory has been developed so far. For instance, effective measure complexity \cite{Grassberger86} uses the block entropies for the distribution of sub-patterns (translation symmetry), \textit{Huberman-Hogg complexity} \cite{Huberman86} counts repetitions of sub-trees in graphs (again, translation symmetry), \textit{permutation complexity} \cite{Bandt02} analyses invariance under permutations and \textit{holographic complexity} considers spacetime symmetries \cite{Brown16, Bernamonti19}. The well-known \textit{algorithmic complexity} (AC) \cite{Li97} defines the complexity of an object by the length of the smallest computer program capable of reproducing the object, which is a way of measuring the compressibility of the object, involving the consideration of all its symmetries. AC has been shown, for instance, to correlate with the order of the automorphism group of graphs \cite{Zenil14}.  

\begin{figure*}
	\centering
	\includegraphics[width=14cm]{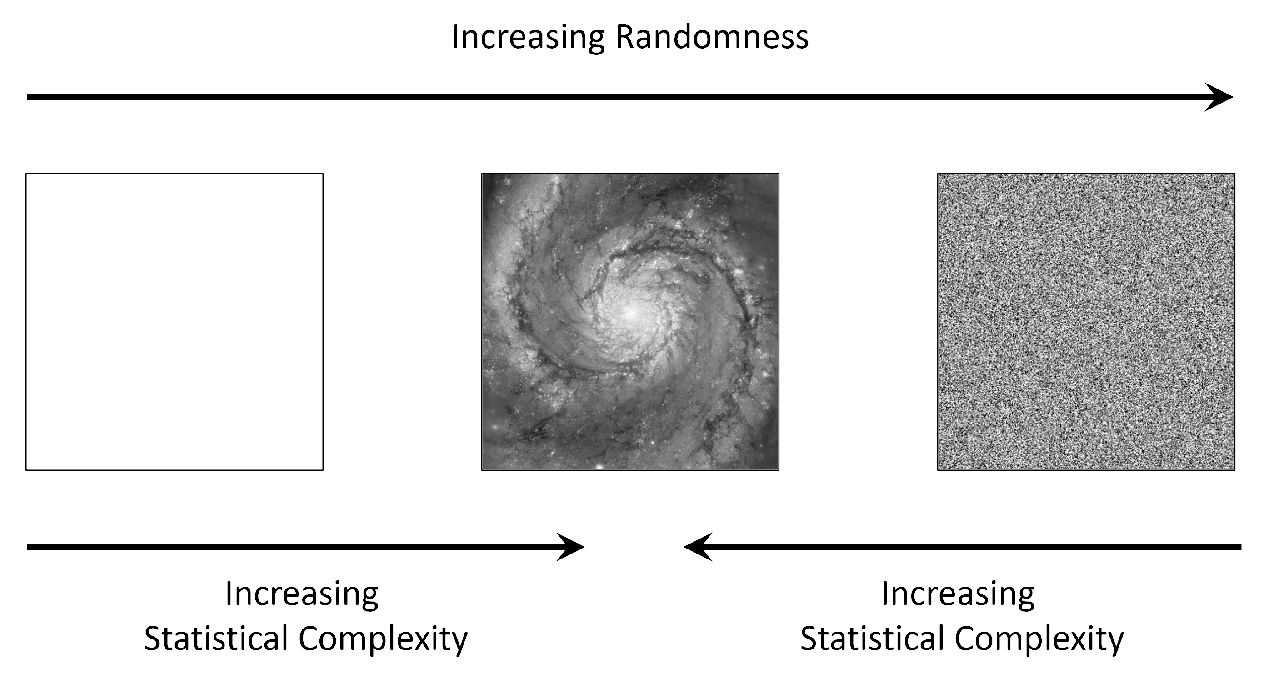}
	\caption{Qualitative comparison between the two main groups of complexity measures on how they would classify three different black and white images. Middle image: Whirlpool Galaxy (desaturated by the author), courtesy of NASA and The Hubble Heritage Team STScI/AURA.}
	\label{figure:cgroups}
\end{figure*}

The great majority of complexity measures can be classified in two groups: those associating complexity to randomness, called here randomness measures, which include AC \cite{Li97} and Lempel-Ziv complexity (LZ) \cite{Lempel76}, and statistical complexities, which attribute higher complexity to objects that can be understood as being between totally ordered and totally disordered configurations \cite{Huberman86, Grassberger86, Crutchfield89}. The latter has been favored by physicists due to their connection to thermodynamics -- very high or very low temperature systems are usually simpler to describe, while intermediate ones are theoretically more challenging. Equivalently, both homogeneous and random structures have simple properties, with more interesting behaviors found in between. Figure \ref{figure:cgroups} illustrates the difference between these two groups using black and white (binary) images as examples. The blank image to the left is considered structurally simple for all kinds of measures due to its exact homogeneity in terms of the distribution of colored pixels. The two groups also agree on classifying the middle image, a desaturated picture of a galaxy, as more complex than the blank one. However, the rightmost image, which is a realization of a square grid where every pixel has the same probability of being black or white, is considered the most complex for those measuring randomness, while it is simple for statistical complexities.

Other relevant properties have yet been proposed. Emmert \cite{Emmert10} argued that complexity is relative rather than absolute and introduced the \textit{statistic complexity}, which provides a relative value of complexity between two objects. A common heuristics used in the process of designing a complexity measure is to first establish what kind of objects attain its minimum or maximum value according to the intended application, which corroborates Emmert's analysis. A well-known example is the disequilibrium-based complexity of L\'opez-Ruiz \textit{et al.} \cite{Lopez95}, which is designed to be 0 when the probability distribution for the states of an object is either uniform or trivial (i.e, has only one possible state) with other distributions having higher values. A related point is raised by Adami \cite{Adami00}, who argues that complexity depends on the meaning that the environment imbues to the object, in the sense that the ensemble of objects to which it should be compared needs to be pre-specified.

In this work, it is proposed that a symmetry-based principle, named here statistical asymmetry, can be used as an underlying principle connecting the existing ideas and measures of complexity. The main steps of a general methodology to find complexity measures based on this principle is outlined and, using the example of binary objects, it is shown how it can reveal the connection between randomness measures and statistical complexities. 

The principle of statistical asymmetry is explained in details in section \ref{section:SA}, where the main steps of the general methodology are introduced. In section \ref{section:CaR}, it is shown how the proposed methodology can be used to design a measure of complexity capable of detecting, within limits to be explained, the level of randomness of binary objects. Then, in section \ref{section:SC}, the same methodology is used again with binary objects, this time to derive a statistical complexity measure which is then applied to obtain the complexity of the configurations of a statistical ferromagnetic spin model (Ising model). A discussion about the significance of the obtained results and the next steps to develop further the framework is presented in section \ref{section:Conclusions}.

\section{Statistical Asymmetry}
\label{section:SA}

Consider an object the description of which can be given by a mathematical quantity $\sigma$. Call $\TS$ the set of transformations acting on $\sigma$, with the specific requirement that the identity transformation, denoted by $I$, is an element of this set. For simplicity, $\TS$ is considered to be finite and discrete, although this assumption can be dropped by appropriately adapting the theory's expressions. Call $M=|\TS|$ the cardinality of $\TS$ and $T\sigma$ the image of $\sigma$ under a transformation $T\in\TS$. If $T$ leaves a certain function $f(\sigma)$ invariant, then it is called a \textit{symmetry} of $\sigma$. If every $T\in\TS$ is a symmetry of $\sigma$, then $\sigma$ is said to be (fully) symmetric under the set $\TS$, otherwise it is not-symmetric. Strictly speaking, $\TS$ does not need to be a group, although it usually is.

In order to go beyond a yes/no binary classification (symmetric/not symmetric), a \textit{dissimilarity measure} $d(\sigma,\sigma')$ capable of evaluating how different the object is from its images is needed. For convenience, assume $d(\sigma,\sigma')>0$ and $d(\sigma,\sigma)=0$. Using this measure, the probability density function (PDF) for the dissimilarity values as the object $\sigma$ is acted upon by the elements of $\TS$ can be written as
\begin{equation}
	\probd{x|\sigma,\TS} = \frac1M \sum_{m=1}^M \delta\prs{x-d_m},
	\label{equation:Kdist}
\end{equation}
where $\delta(\cdot)$ is a Dirac delta function, $d_m = d(\sigma, T_m\sigma)$ and $T_m$, $m=1,...,M$,  are the elements of $\TS$. This density encodes statistical information about how asymmetric the object is under $\TS$. For a fully symmetric object under $\TS$, for instance, it should be a delta at zero. If the distance is a discrete variable, a case that will be addressed in the next section, one can then work directly with the appropriate probability mass function (PMF). 

Let us now define the measure of statistical asymmetry
\begin{equation}
	A\prs{\sigma} = \Delta\col{\probd{x|\sigma,\TS}||\mathcal{Q}_d\prs{x|\sigma,\TS}},
\end{equation}  
where $\Delta[P||Q]$ is a distance between probability distributions $P$ and $Q$ and $\mathcal{Q}$ is a reference distribution which contains information about what kind of symmetries and what environments should be used to define what is complexity in a particular context. How exactly the values of $A$ will relate to the object's amount of asymmetry depends on the reference distribution, which will be discussed in more details in the next sections.

Caution has to be taken when applying the above expression due to many subtleties for every choice that can be made. The choice of $\TS$, for instance, can radically change the obtained values and lead to apparently contradictory consequences, as we will discuss when we analyze the case of binary objects. The distance $d$, as another example, might be designed to detect only specific differences, ignoring others. Each one of these choices needs to reflect what properties are important for the application that is being considered. This allows for a better understanding of the role of each of these characteristics and how they contribute to the complexity of an object.

\section{Complexity as Randomness}
\label{section:CaR}

In this and the next section, the expressions presented formerly will be applied to the study of binary objects for simplicity. With due care, extensions to more general objects can be obtained. Consider $\sigma$ to be a binary array with $N$ elements. The focus is on one and two dimensional arrays -- binary sequences and matrices -- but the expressions presented here can be straightforwardly generalized to higher dimensions.   

For binary arrays, a convenient choice for a measure of dissimilarity is the Hamming distance, commonly used in coding theory \cite{Richardson08} to measure the bitwise error of decoding (the dissimilarity between original and decoded messages). It is defined as the number of mismatching symbols between two arrays $\sigma$ and $\sigma'$. Choosing the binary alphabet to be $\prs{-1, +1}$ by convention, the Hamming distance is written as
\begin{equation} 
	d(\sigma, \sigma') =\sum_i \frac{1-\sigma_i\sigma'_i}{2} = \frac{N}2-\frac12\sum_i\sigma_i\sigma'_i,
	\label{equation:HDist}
\end{equation}
where the sum is over all elements of the array. The Hamming distance can assume any integer value between 0 and $N$. Being discrete, one can use its PMF in the formulas below which, by an abuse of notation, will be denoted by the same symbol $\probd{x|\sigma,\TS}$. This should not be a source of confusion in general, as the nature of the variable will indicate whether we use the PDF or PMF in each case. A natural choice for $\Delta$ would be the Kullback-Leibler (KL) divergence, the foundations of which lie in information theoretical principles. If the reference distribution to which we compare ththat of the of dissimilarity values is chosen to be the uniform distribution $\mathcal{Q}_d\prs{x|\sigma,\TS}=1/(N+1)$, which has the maximum entropy for a discrete distribution, this leads to
\begin{equation}
	\begin{split}
		A\prs{\sigma} &= \sum_{i=0}^N \probd{x|\sigma,\TS} \ln \frac{\probd{x|\sigma,\TS}}{1/(N+1)} \\
					  &= \ln (N+1)-\sum_i \probd{x|\sigma,\TS} \ln \probd{x|\sigma,\TS}\\
	              	  &= S[\mathcal{Q}_d\prs{x|\sigma,\TS}]-S[\probd{x|\sigma,\TS}],
	\end{split}
\end{equation}  
where $S[\cdot]$ is the (discrete) Shannon's entropy. The first term is a constant for all objects, meaning that different objects will only be differentiated by the entropy of their particular dissimilarity distribution. The latter has the property that, if the object is completely symmetric under the transformation set, its value is zero (as $\probd{x=0|\sigma,\TS}=1$ and zero otherwise). On the other hand, if the Hamming distance assumes a different value for all images of the object, then one can argue that it is the least symmetric possible under the given transformation set. This suggests that, ignoring the first constant term and taking the complexity measure as simply
\begin{equation}
	B\prs{\sigma} = S[\probd{x|\sigma,\TS}],
\end{equation}  
associating complexity to randomness through the object's amount of statistical asymmetry. 

The measure $B$ as defined above cannot give a perfect characterization of randomness, but only one that is relative to its compressibility under the set $\TS$. The characterization of full randomness, which becomes associated with the total lack of symmetry or compressibility, is subtle and better captured by AC. Theoretically, it is described by the Martin-L\"of theory \cite{MartinLoef66}, although it becomes rigorous only for infinite sequences of symbols. Nevertheless, there are Martin-L\"of tests to assess the extent to which finite sequences can be given an intuitive classification as random. The differences in $B$ and AC are due to the fact that the former does not take into consideration all possible patterns, or symmetries, in the object, but only a subset of them. Yet, $B$ can be very successful in capturing aspects of randomness under its limitations, as it is shown in the following. 

In order to illustrate the above considerations, let us apply the measure $B$ to calculate the complexity of a succession of binary sequences of length $N$ with periodic boundary conditions (PBCs) and $\TS$ as the set of $n$-steps rotations to the right, i.e., rotations that take the element $i$ of the sequence to position $i+n$. Because PBCs imply $i+N =i$, they effectively turn the one-dimensional array into a circle. The succession of sequences is obtained by the following stochastic process, which we call `erosion' for simplicity. The first sequence contains $N$ elements, all of them are the same and assumed to be -1. At each discrete time step, an element of the sequence is chosen uniformly randomly and is multiplied by -1 with probability 1/2. Asymptotically, this transforms a homogeneous sequence, the complexity of which is clearly $B=0$, to a sequence that should approach what one would expect to be, intuitively, a random one. 

\begin{figure*}
	\centering
	\includegraphics[width=18cm]{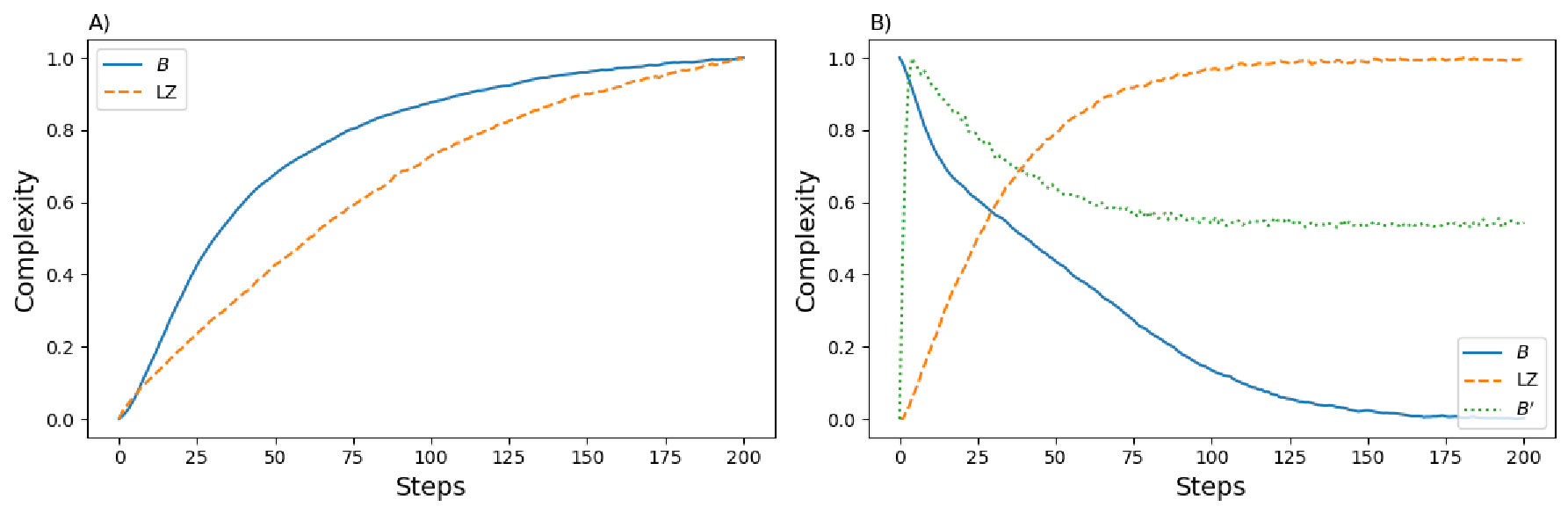}
	\caption{Comparisons between (A) the average evolution of $B$ and LZ for binary sequences produced in the stochastic erosion process and (B) the same two quantities together with $B'$ for the stochastic mixing process.}
	\label{figure:bvslz}
\end{figure*}

In figure \ref{figure:bvslz}A, the full line shows the values obtained by calculating $B$ for a binary array of length $N=200$, run for 200 steps and averaged over 200 realizations of the process of erosion. The dashed line is the value obtained by calculating the LZ complexity for the same sequences. Both quantities are normalized by subtracting their minimum values and dividing by their maximum for better comparison. LZ is considered to be a practical approximation for AC, which is uncomputable, and is the measure of compressibility that serves as the basis for the compression of GIF files. 

Like LZ, $B$ increases monotonically as the sequence becomes more random, which shows that $B$ is indeed capable of measuring its increasing randomness. On the plot of figure \ref{figure:bvslz}B however, the behavior of both measures is quite different. The plot compares again both values for another succession of sequences. The process, which we call `mixing', now starts with a sequence the first half of which is composed only of -1's and the second half of 1's. At each discrete step, a pair of elements is chosen uniformly randomly and exchanged. While LZ increases as before, $B$ now decreases until it shows a tendency to stabilize. The comparison shows that, for the chosen transformations, $B$ cannot detect randomness as LZ does. The reason for that is the choice of $\TS$, which restricts which symmetries $B$ can detect, while LZ has, in principle, no restrictions at all. In the case of these sequences, LZ can detect that the initial sequence is invariant by the exchange of blocks with the same value, something that $B$ cannot by construction. It is worth noticing that $B$ only converges to a zero value because we are subtracting its minimum. Such a minimum, however, is the same as the maximum of the previous process. On the other hand, such anti-correlated behavior is evidence that it is possible to use the same principle to obtain a measure with properties similar to those of statistical complexities, what is shown in the next section.

To illustrate the fact that a choice of transformations more similar to what is done in the LZ procedure would lead to closer values of the two measures, the same figure \ref{figure:bvslz} shows a plot of the quantity $B'$, which only differs from $B$ in the choice of $\TS$. As it would be impractical to check the sequence under a set of transformations composed by all possible permutations of blocks of varying sizes, $B'$ only uses blocks of size 2. Even though, for a sequence of 200 symbols, this would lead to 100! permutations of blocks. Instead, a statistical approach is adopted and 200 random permutations are compared to the original sequence at each step. The result is now closer to LZ than that of $B$ as expected. 

\section{Statistical Complexity}
\label{section:SC}

Statistical asymmetry can also be used to derive a measure of statistical complexity, what is shown in the following. The use of the Hamming distance and Shannon's entropy in the randomness measure $B$ indicates that the expression is ``looking'' at all the microscopic differences in the initial object and the transformed one, where `microscopic' means that every mismatched symbol contributes equally, independently of the size of the object. The key insight to obtain a different measure can be understood through a thought experiment. Consider a square grid where cells can be either white or black. If each cell has a probability $p$ of being black, then the greater the value of $p$ the higher the density of black cells on a random realization of the grid. If every cell is discernible, one can differentiate two realizations by the different positions of the black (or white) ones. Now, consider walking away from the grid. As the distance increases, the pattern of black cells becomes increasingly blurred. At some point, different random realizations for the the same probability $p$ become indistinguishable with all random patterns looking like homogeneous gray surfaces differing only by the hues of grey, which reflect the average density of black cells. In other words, macroscopically, random distributions of black cells are equivalent to homogeneous ones of different densities. The difference between statistical complexities and randomness measures then becomes an issue of scale, or coarse graining. There is a clear relation to renormalization group theory here, but we will not explore it at this point in order not to deviate from our main objective. 

To translate the above ideas mathematically, first notice that the maximum value of the Hamming distance depends on $N$ and is therefore sensitive to the scale of the object. In order to look at the object from a distance, it is necessary to work with a normalized version of it where $d(\sigma, \sigma')$ is the Hamming distance divided by $N$, which constrains it to the interval $[0,1]$. This will also allow for a meaningful comparison of objects of different size and will be important in the limit of large systems, a common case of interest.  

Symmetry considerations lead us to expect that in random sequences (in the Martin-L\"of sense) the probability distribution of the (normalized) dissimilarities should peak with decreasing spread around a certain typical average value. The terms of the sum in the equation for the Hamming distance should behave as independent random variables and obey the central limit theorem (CLT) for large structures. Based on this argument, let us choose $\mathcal{Q}_d\prs{x|\sigma,\TS} =\mathcal{N}\prs{x|\bar{d}, s_d^2 }$, a normal distribution with the same mean $\bar{d}$ and variance $s_d^2$ as $\probd{x|\sigma,\TS}$, guaranteeing that both homogeneous and random structures have low complexity compared to others. 

Again, an intuitive choice for the $\Delta$ would be the KL divergence. In practice though, this leads to numerical difficulties. As the distribution of dissimilarities is not expected to be analytically obtainable (except in very particular cases), the logarithms within KL become too sensitive to the statistical variations resulting from its numerical estimate. A more stable choice is a $L^2$ distance between the two densities. It is convenient also to use a kernel density estimate $\hat{P}_d$ to approximate $P_d$ with a Gaussian kernel \cite{Silverman98}
\begin{equation}
	\hat{P}_d (x)= \frac1M \sum_{i=1}^M \mathcal{N}\prs{x|d_m, h^2},
\end{equation}
where Silverman's rule is used for the standard deviation of the kernels $h=\prs{\frac{4s_d^5}{3M}}^{1/5}$. This way, the integral can be solved analytically to give 
\begin{equation}
	\begin{split}
		A\prs{\sigma} &= \int_{-\infty}^\infty dx \col{\hat{P}_d (x)-\mathcal{N}\prs{x|\bar{d}, s_d^2}}^2 \\
		              &= \frac1{M^2}\sum_{m,n=1}^M \frac1{2h\sqrt{\pi}} e^{-\frac{(d_m-d_n)^2}{4h^2}}\\
		              &  \quad-\frac{2}{M}\frac1{\sqrt{2\pi(h^2+s_d^2)}}e^{-\frac{(d_m-\bar{d})^2}{2(h^2+s_d^2)}}
		              +\frac1{2h\sqrt{\pi}}. 
	\end{split}
\end{equation} 

One of the most versatile model systems in statistical physics, with applications in areas from AI to economics, is the Ising model, originally a simplified description of a magnetic material. The `material' is represented by a lattice with binary variables at each site. In the ferromagnetic case, there is a tendency of neighbors in the lattice to have the same value. In one dimension, where the lattice becomes a line, the behavior is almost trivial, but for a two-dimensional square lattice, the model is capable of reproducing the continuous phase transition that characterizes the change between ferromagnetic and paramagnetic materials at the critical temperature. Exact analytical solutions characterizing the relevant quantities of these two cases are known \cite{Baxter89, Onsager44}. 

\begin{figure}
	\centering
	\includegraphics[width=8.5cm]{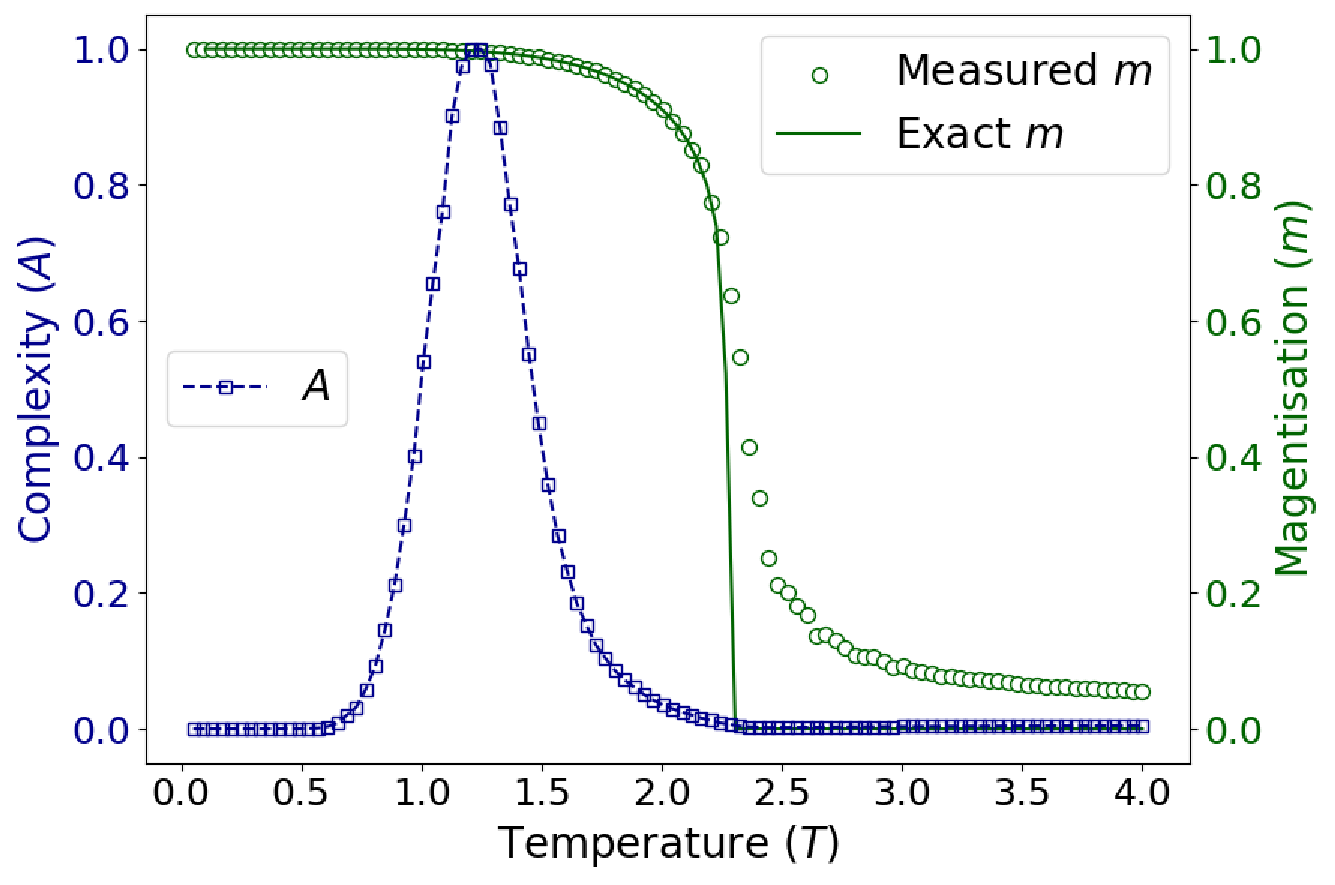}
	\caption{Complexity measure $A$, derived from the statistical asymmetry principle, as a function of the temperature for the 2D Ising model on a square lattice (dashed, squares). The magnetization per spin is plotted in the same graph both for the used sequences (circles) and as the exact solution for an infinite lattice (full line).}
	\label{figure:2dimtrans}
\end{figure}

Fig. \ref{figure:2dimtrans} shows the value of $A$ (normalized by its maximum value) for the 2D ferromagnetic Ising model on a square lattice with 900 spins averaged over 5000 configurations per temperature (the Boltzmann constant and the spin coupling are both set to 1). The magnetization per spin for these configurations and the exact result for an infinite lattice are plotted together as references to the phase transition. 

The qualitative behavior of the complexity measure $A$ is what one would expect from a statistical complexity. The position and spread of the peak are functions of both the size of the lattice and the set of transformations, as it was discussed briefly in the case of $B$. The position of the peak indicates a temperature where the configurations seem to lose (statistically) their translational symmetry. It is unlikely that it has any physical consequence as the choice of the transformation group should be considered further to reflect relevant features of the system. Nevertheless, it is able to capture the behavior that was the one sought by the statistical complexity measures. 

\section{Conclusions}
\label{section:Conclusions}

The ability of the statistical asymmetry principle as introduced in this work to allow the design of complexity measures capable of reproducing the behavior of both randomness measures and statistical complexities from fundamental principles is the main result of this work. It shows that statistical asymmetry makes explicit the missing connection between symmetry and complexity.

We conjecture that he steps outlined here should allow to reproduce a large part, if not all, existing complexities by judiciously choosing the appropriate distances and symmetries relevant in each application. The examples that have been analyzed, although not constituting a definitive proof, show strong evidence in support of this conjecture. The main challenge is to find a systematic way of making the relevant choices for (i) the transformation set, (ii) the similarity measure and the (iii) probability distance which reflect the relevant features of complexity that need to be captured in each different application. It is interesting that a similar problem has been identified in connection to pattern recognition in neural networks, with a methodology to deal for identifying symmetries in finite objects being developed for that application \cite{Minsky88}. Whether and how such framework can contribute to elucidate issues of the one presented here is a question that is currently being pursued.

The link unveiled in this work opens up several lines of investigation. An first task would be to list the currently used complexities and find the appropriate choices to reproduce their results, possibly uncovering interesting new ways of looking at the concept. Applying the measures introduced here to the practical problems listed in the introduction of this article should provide additional insights on the relative importance of transformation sets and distances in different areas and scenarios. Such exciting possibilities are currently being investigated.

\section*{Acknowledgments}
The author would like to acknowledge insightful discussions with Dr Juan Neirotti, Prof David Saad, Dr Elizabeth Wanner and Dr Jens Christian Claussen.

\bibliographystyle{apsrev4-2}      
\bibliography{complex}   

\end{document}